\documentclass[10pt]{iopart}
\usepackage{iopams}
\usepackage{graphicx}
\usepackage{cite}
\usepackage{color}
\usepackage{multirow,dcolumn}
\begin{document}
\title[All-electric Experiments for Piezoelectric Phononic Plates]{Piezoelectric Phononic Plates: Retrieving the Frequency Band Structure via All-electric Experiments.}
\author{F H Chikh-Bled, N Kherraz\footnote{Present address: Univ. Lille, CNRS, Centrale Lille, ISEN, Univ. Valenciennes, UMR 8520-IEMN, F-59000 Lille, France}, R Sainidou, P Rembert and B Morvan}

\address{Laboratoire Ondes et Milieux Complexes UMR CNRS 6294, UNIHAVRE, Normandie University, 75 rue Bellot, 76600 Le Havre, France}

\ead{bruno.morvan@univ-lehavre.fr}

\vspace{10pt}
\begin{indented}
\item[]\today
\end{indented}

\begin{abstract}
We propose an experimental technique based on all-electric measurements to retrieve the frequency response of a one-dimensional piezoelectric phononic crystal plate, structured periodically with millimeter-scaled metallic strips on its two surfaces. The metallic electrodes, used for the excitation of Lamb-like guided modes in the plate, ensure at the same time control of their dispersion by means of externally loaded electric circuits that offer non-destructive tunability in the frequency response of these structures. Our results, in very good agreement with finite-element numerical predictions, reveal interesting symmetry aspects that are employed to analyze the frequency band structure of such crystals. More importantly, Lamb-like guided modes interact with electric-resonant bands induced by inductance loads on the plate, whose form and symmetry are discussed and analyzed in depth, showing unprecedented dispersion characteristics.
\end{abstract}

\vspace{2pc}

\maketitle



\section{Introduction}

Phononics has shown a significant growth in the last two decades, due to the ability of periodic structures to exhibit unusual dispersion properties not encountered in ordinary materials, triggered by the periodicity~\cite{deymier,khelif}. Recently, new modern topics such as heat control~\cite{yu}, topological~\cite{zyang,miniaci} and non-reciprocal~\cite{maznev,trainiti,croenne} wave phenomena, etc. have imbued this area, while in parallel a noticeable effort has been focused on the combination of phononics with active materials in order to provide non destructive control of elastic waves within these structures. Among several possibilities, the use of piezoelectric materials constitutes an important source of inspiration for the design and realization of intelligent active devices based on their ability to combine electric circuits that offer ease in non-destructive, electric command of the dispersion properties of these structures.

Most of the efforts have been focused on phononic crystal structures containing piezoelectric materials, especially in the form of thin patch arrays at the surface of, and/or thin inserts within otherwise homogeneous materials~\cite{thorp,baz,spadoni,casadei,yychen,degraeve1,degraeve2,sugino} in order to ensure their coupling to external circuits that control electrically the frequency response of the whole structure. The band structure of these systems becomes tunable, its form depending on the nature of the electrical shunt connected on the piezoelectric elements. The targeted application is mainly the vibration mitigation using tailored frequency band gaps~\cite{spadoni,hagood,beck,banerjee}.

Independently, use of piezoelectric materials in phononic crystals goes back to the 2000s~\cite{ttwu,laude,bench,hladky}. Motivated by the study of surface acoustic waves, some of these works involve a piezoceramic plate as the host matrix for two-dimensional holey arrays. Interestingly, such configurations integrate on the same piezoelectric plate the phononic crystal and two interdigital transducers used to generate and receive the guided waves; they do not offer however the possibility of tuning the frequency band structure of the crystal. One-dimensional (1D) phononic crystals including homogeneous piezoelectric plates with corrugated surfaces~\cite{huang1,huang2} have been also considered.

To the best of our knowledge, only a few works propose piezoelectric materials as host matrices for the design and realization of active phononic devices. This choice take advantage of the stronger electromechanical coupling that occurs within the piezoelectric material, usually in the form of plate on the surfaces of which a 1D array of metallic electrode strips are designed~\cite{kherraz1,kherraz2,kherraz3,vasseurc}. In this case, the propagation medium is considered elastically homogeneous (no discontinuities are imposed from a mechanical point of view, since the electrodes are supposed to be of negligible thickness) and the periodic nature of the structure relates exclusively to the electric boundary conditions (EBCs) applied to the periodically distributed electrodes, deposited on the surfaces of the plate. Thus the electric impedance loads connected on the electrodes can be used via the electromechanical coupling to actively control the propagation of Lamb-like guided waves within the plate. For instance, an inductive shunt in association with the inherent capacitive nature of the piezoelectric material will operate as an equivalent electric resonant circuit that produces avoided crossings and, under certain conditions, opening up of hybridization gaps. Though such hybrid modes originating from the admixture of localized electric resonance with Lamb-like modes are predicted theoretically by the authors in previous papers~\cite{kherraz2,kherraz3} for various electric-load configurations, up to now no experimental investigation of their dispersion properties has been realized.

In this paper, we present an alternative method based on all-electric measurements to retrieve the frequency band structure of a one-dimensional piezoelectric phononic crystal plate coupled with external circuits inducing electric impedance loads on the structure. Precisely, we make use of both the electric potential and impedance recorded along the structure to analyze the Lamb-like guided modes generated in the plate, combined to localized resonance modes induced by external inductance loads. The symmetry of the frequency bands is also discussed in some detail in relation to the symmetry of the unit cell.

The paper is structured as follows. First, in~\sref{sec:methods} we outline the methods used in the experiments and describe the structures under study. In~\sref{sec:results}, we employ three model cases to develop our methodology, discussing in detail physical aspects and analyzing in terms of symmetry the several modes in the dispersion plots. Finally,~\sref{sec:conclusion} concludes the paper.

\section{Methods}\label{sec:methods}

\subsection{Piezoelectric phononic crystal plate}\label{sec:pc}

\begin{figure}
\centering
\includegraphics[width=8.2cm]{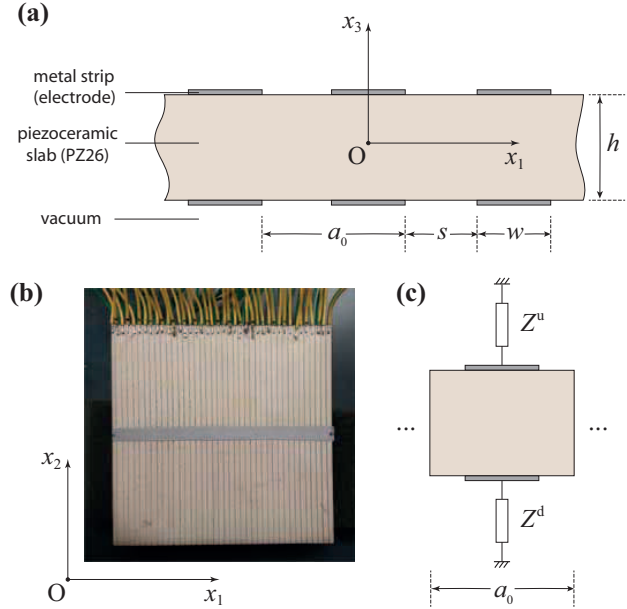}
\caption{\pt(a) Schematic representation of the one-dimensional piezoelectric crystal, extended to infinity along $x_1$ and $x_2$ directions, which coincide with the transversely isotropic plane of the piezoceramic material, poled across its thickness along the $x_3$ symmetry axis. \pt(b) An image of the fabricated sample with finite dimensions ($80~\mathrm{mm}\times80~\mathrm{mm}\times2.2~\mathrm{mm}$), used in the experiments, with the electrodes connected at the edges of the metallic strips (width $w=1.7~\mathrm{mm}$ and separation gap $s=0.3~\mathrm{mm}$) aligned along $x_2$ direction. The structure consists of $N=40$ elementary blocks of length $a_0=2~\mathrm{mm}$. In \pt(c) an elementary block of the structure loaded with electric circuits of impedance $Z^{\mathrm{u}}$ (upper plate side) and $Z^{\mathrm{d}}$ (lower plate side).}\label{fig1}
\end{figure}
We consider a 1D piezoelectric phononic crystal plate of lattice constant $a$. A schematic representation of the structure is depicted in \fref{fig1}\pt(a). The crystal consists of an array of parallel metallic strips (electrodes) of width $w$ and separated by a distance $s$, placed symmetrically (i.e., face-to-face) on both sides of an otherwise homogeneous piezoelectric plate of thickness $h$ and are assumed to be of negligible thickness. The sample used in the experiments is shown in \fref{fig1}\pt(b). The square piezoelectric plate (edge length of $80~\mathrm{mm}$, thickness $h=2.21~\mathrm{mm}$) is made of PZ26, polarized along $x_3$-axis, and a silver thin film, $15~\mathrm{\mu m}$-thick, is deposited on the full area of both surfaces of the plate. Parallel grooves of average width $s=0.3~\mathrm{mm}$ were machined to create the array of strips whose average width is $w=1.7~\mathrm{mm}$. Wires welded on these rectangular electrodes ensure all electric connections used in the experiments. Each pair of electrodes (up and down side of the plate) can be connected to electric circuits of characteristic impedance $Z^{\mathrm{u}}$ and $Z^{\mathrm{d}}$ [see \fref{fig1}\pt(c)], thus implying that the elementary blocks of length $a_0=w+s$, may have identical or different EBCs throughout the structure. This offers the possibility of tuning the periodicity in a non-destructive, active, manner. In the simplest case, when the same EBCs are applied to every elementary block, the lattice parameter is $a=a_0$; in the general case, $a$ may be a multiple of $a_0$.

\begin{figure*}[t]
\centering
\includegraphics[width=12cm]{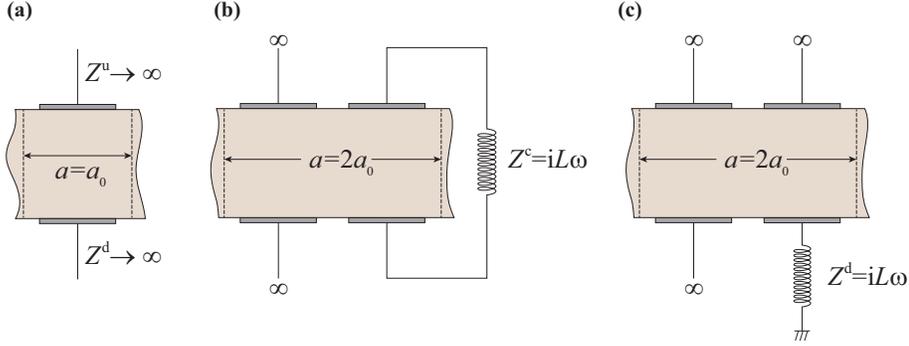}
\caption{The unit cell of the piezoelectric phononic crystal of lattice constant $a$, for the three configurations under study: \pt(a) \emph{system I}: uniform floating-potential EBCs for all electrodes ($a=a_0$) applied on both sides ($Z^{\mathrm{u}}=Z^{\mathrm{d}}\rightarrow \infty$), \pt(b) \emph{system II}: alternated EBCs ($a=2a_0$) with floating potential applied to the first pair of electrodes and an inductance load ($Z^{\mathrm{p}}=\mathrm{i}L \omega$) applied to the second pair of electrodes, and, \pt(c) \emph{system III}: alternated EBCs ($a=2a_0$) with floating potential applied to the first pair of electrodes and on the upper-side electrode of the second pair, while the corresponding lower-side electrode is inductance-loaded ($Z^{\mathrm{d}}=\mathrm{i}L \omega$).}\label{fig2}
\end{figure*}
We shall be concerned with three distinct configurations, schematically represented in \fref{fig2}: (i) the case of identical EBCs, precisely all electrodes having floating potential, i.e., $a=a_0=2~\mathrm{mm}$ and $Z^{\mathrm{u}}=Z^{\mathrm{d}}\rightarrow \infty$; we shall refer to it as \emph{system I}, (ii) alternated EBCs in a unit cell spanning over two elementary blocks, i.e., $a=2a_0=4~\mathrm{mm}$, the first having floating potential ($Z^{\mathrm{u}}=Z^{\mathrm{d}}\rightarrow \infty$), the second loaded with an inductance $L$ (of impedance $Z^{\mathrm{p}}=\mathrm{i}L \omega$) connected in parallel with the piezoelectric plate; we shall refer to it as \emph{system II}, and, (iii) alternated EBCs in a unit cell spanning over two elementary blocks, i.e., $a=2a_0=4~\mathrm{mm}$, the first having floating potential ($Z^{\mathrm{u}}=Z^{\mathrm{d}}\rightarrow \infty$), the second loaded with an inductance $L$ connected in series with the lower-side of the piezoelectric plate ($Z^{\mathrm{d}}=\mathrm{i}L \omega$), while the upper-side has floating-potential conditions; we shall refer to it as \emph{system III}.

\subsection{Experimental Setup}\label{sec:setup}

\begin{figure}
\centering
\includegraphics[width=8.2cm]{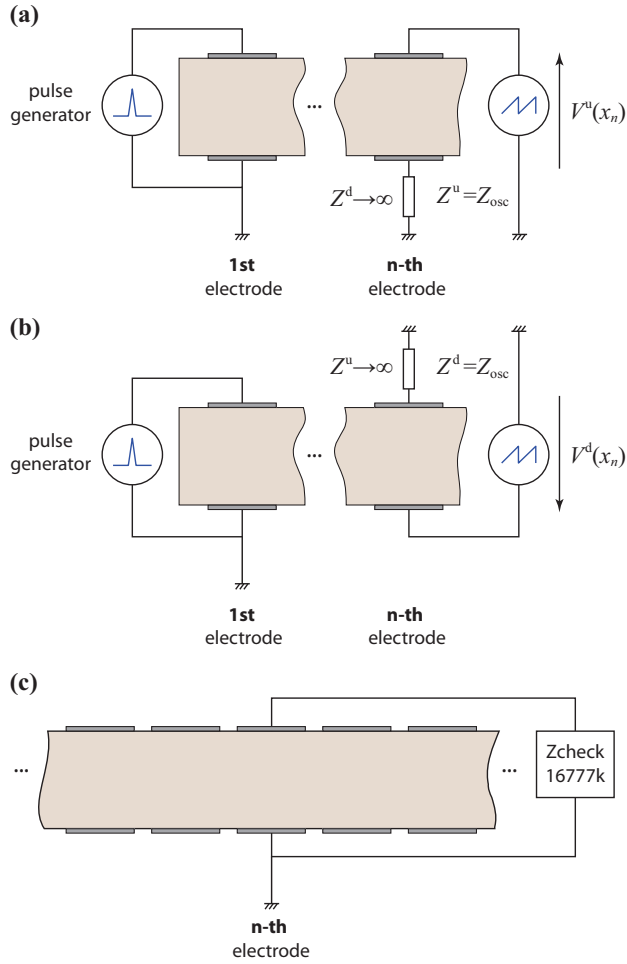}
\caption{Experimental configurations for the recording of the electric potential in \pt(a) the upper, and \pt(b) the lower side of the plate, along the direction of the array, at positions $x_n$ associated to the $n$-th pair of electrodes, with a pulse excitation applied at the first electrode. In \pt(c), we represent schematically the experiment setup for the electric impedance measurements at a position $x_n$ associated to the $n$-th pair of electrodes, via an impedance analyzer, which after excitation with a sinusoidal voltage collects the electric response at this position.}\label{fig3}
\end{figure}
Two main experimental setups are considered, respectively associated to electric potential, and, electric impedance measurements. In the first case, we expect that, after excitation of the piezoelectric plate, any deformations will be manifested as electric potential variations on the electrodes through the electromechanical coupling that takes place within the piezoelectric material. A pulse generator (Panametrics model 5058PR) allows to excite guided waves propagating in the plate: a $200\mathrm{V}$-amplitude and $0.1\mathrm{\mu s}$-width pulse is applied to the first pair of electrodes at the one edge of the piezoelectric plate (let's say, position $n=1$), the electromechanical coupling ensuring a broadband excitation of these waves. The electric potential is recorded with a $10$-bit quantification, at each electrode position, on the upper or lower side of the plate, with the help of a digital oscilloscope (LeCroy HRO66ZI WaveRunner), the ground of the pulse generator being taken as reference [see \fref{fig3}\pt(a), \pt(b)]; the time window chosen is typically set to $250~\mu \mathrm{s}$, with a sampling period equal to $5~\mathrm{ns}$, allowing to observe a few forward and backward traveling waves after reflections at the edges of the plate. Each measured signal is then averaged $30$ times to improve the signal-to-noise ratio. The above settings ensure a sufficiently high Nyquist frequency and accurate resolution of the spectra obtained through fast Fourier transforms (FFT).

Electric impedance measurements are also performed using a Zcheck 16777k impedance analyzer. This device evaluates the complex admittance $Y$ and/or impedance $Z$ of an electric dipole, after excitation with a sinusoidal voltage applied at the one pole, the other being connected to the ground; the electric current intensity is then measured to deduce the admittance $Y$ in a frequency range extending from $10~\mathrm{Hz}$ to $16~\mathrm{MHz}$. We employ this technique to record the $Y_n$ and $Z_n$ spectra of the phononic crystal piezoelectric plate at different positions $x_n=(n-1)a_0$ along the periodically structured dimension $x_1$, by connecting the impedance analyzer at the $n$-th pair of electrodes, as shown in \fref{fig3}\pt(c); the obtained spectra measured in the frequency range $[10~\mathrm{kHz},1.5~\mathrm{MHz}]$ with a resolution of $50~\mathrm{Hz}$ include the electric response of the whole structure in the applied voltage excitation, carrying a rich information about the guided modes of the plate.

\section{Results and Discussion}\label{sec:results}

\subsection{Model system: uniform floating-potential EBCs}

To begin we consider the case of \emph{system I} consisting of a piezoelectric plate as described in~\sref{sec:pc} on which uniform EBCs are applied, precisely all electrodes on both sides of the plate have identical, floating-potential EBCs. This system is the simplest and most symmetric configuration and offers a model case to explore and understand the basic behavior of the piezoelectric phononic crystal, before any inductance loads are incorporated. We shall use it to present the experimental tools and measurements that we employ to retrieve the dispersive frequency response of our crystal. Two main methodological approaches are used: the electric potential measurements and the electric impedance measurements, both recorded at a pair of electrodes along the finite crystal's periodicity.

\subsubsection{Electric-potential analysis}

\begin{figure}
\centering
\includegraphics[width=8.2cm]{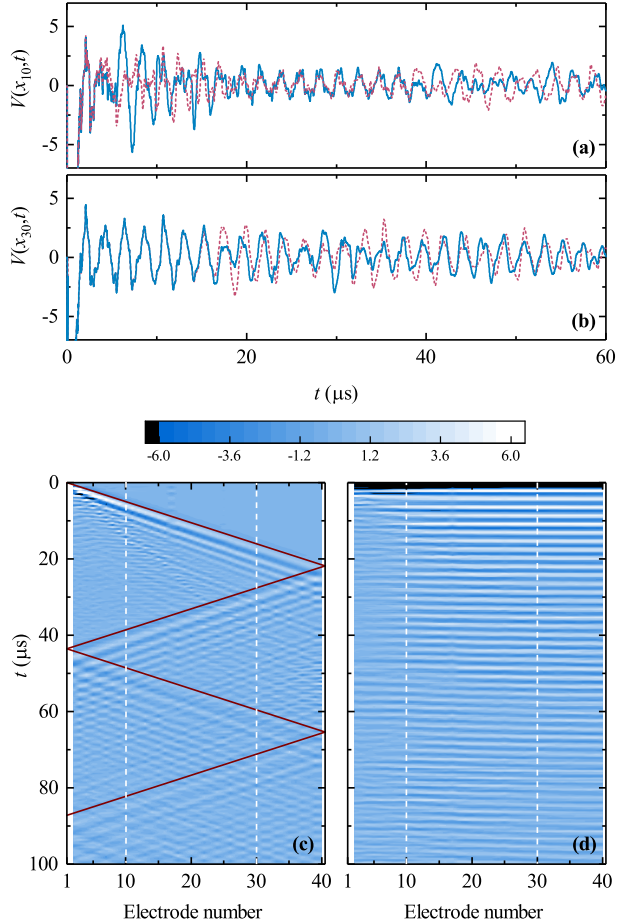}
\caption{Evolution of the upper-side, $V^\mathrm{u}(x_n,t)$, (solid blue line) and lower-side, $V^\mathrm{d}(x_n,t)$, (dashed red line) electric-potential signals, recorded at \pt(a) $n=10$, and, \pt(b) $n=30$ electrode position. In \pt(c) and \pt(d), we show, respectively, the difference, $V^\mathrm{u}(x_n,t)-V^\mathrm{d}(x_n,t)$, and, the sum, $V^\mathrm{u}(x_n,t)+V^\mathrm{d}(x_n,t)$, of the electric-potential signals, recorded along the direction $x_1$ of the piezoelectric phononic crystal plate; the white dashed lines denote the $n=10$ and $n=30$ electrodes, while the solid line in \pt(c) represents the long-wavelength $S_0$ propagating mode with velocity close to that of the corresponding homogeneous metallized plate.}\label{fig4}
\end{figure}
At a first stage, the electric potential is measured at every strip, on both up and down sides of the phononic plate, separately: the respective values, $V^{\mathrm{u}}(x_n,t)$ and $V^{\mathrm{d}}(x_n,t)$, are collected at the electrode positions $x_n=(n-1)a_0$, $n=2,\ldots,40$. One observes a more or less similar form for the signals recorded along $x_1$-direction on both sides of the plate, which are mainly characterized by the presence of two wave fronts: the first propagates with a dominant-wave velocity estimate of $3625~\mathrm{m~s^{-1}}$, close to the long-wavelength effective medium velocity, $c_{\mathrm{eff}}$, of the $S_0$ mode of the corresponding homogeneous piezoceramic plate covered with metallized surfaces on both sides (at $x_3=\pm\frac{h}{2}$)~\cite{ceff}; the second one with a much higher velocity propagation, recognized by the formation of wave fronts, almost parallel to $x_2$ direction. Comparison of up-side and down-side signals at $n=10$ (see \fref{fig4}\pt(a)) reveals some intervals for which $V^{\mathrm{u}}$ and $V^{\mathrm{d}}$ have an opposite phase, precisely for $5~\mu \mathrm{s} \lesssim t \lesssim 10~\mu \mathrm{s}$ and for $t \gtrsim 38~\mu \mathrm{s}$ that coincide, respectively, with the arrivals of the incident and reflected at the end of the plate $S_0$-like mode. Comparison of these signals at $n=30$ (see \fref{fig4}\pt(b)), far away from the excitation point located at $n=1$, clearly shows that for $t \lesssim 17~\mu \mathrm{s}$, for which the $S_0$-like signal is not yet arrived, the electric potentials $V^{\mathrm{u}}$ and $V^{\mathrm{d}}$ are in phase. The above suggest that the $S_0$-like mode is manifested as a symmetric part in the potential, while the fast wave manifests itself as an antisymmetric part in the potential recorded at each side of the plate. The operation $V^{\mathrm{u}}\pm V^{\mathrm{d}}$ should thus decompose the image of the signals $V^{\mathrm{\nu}}(x_n,t)$ into two quasi-independent wave components, of the slow ($S_0$-like) and fast modes. In \fref{fig4}\pt(c) and \pt(d) we plot the difference and sum of the up- and down-side signals, that confirm, respectively, the decomposition into a symmetric part dominated by an easily recognizable $S_0$-like component, and an antisymmetric part dominated by a fast wave component, whose physical origin will become clear later.

\begin{figure}[t!]
\centering
\includegraphics[width=8.2cm]{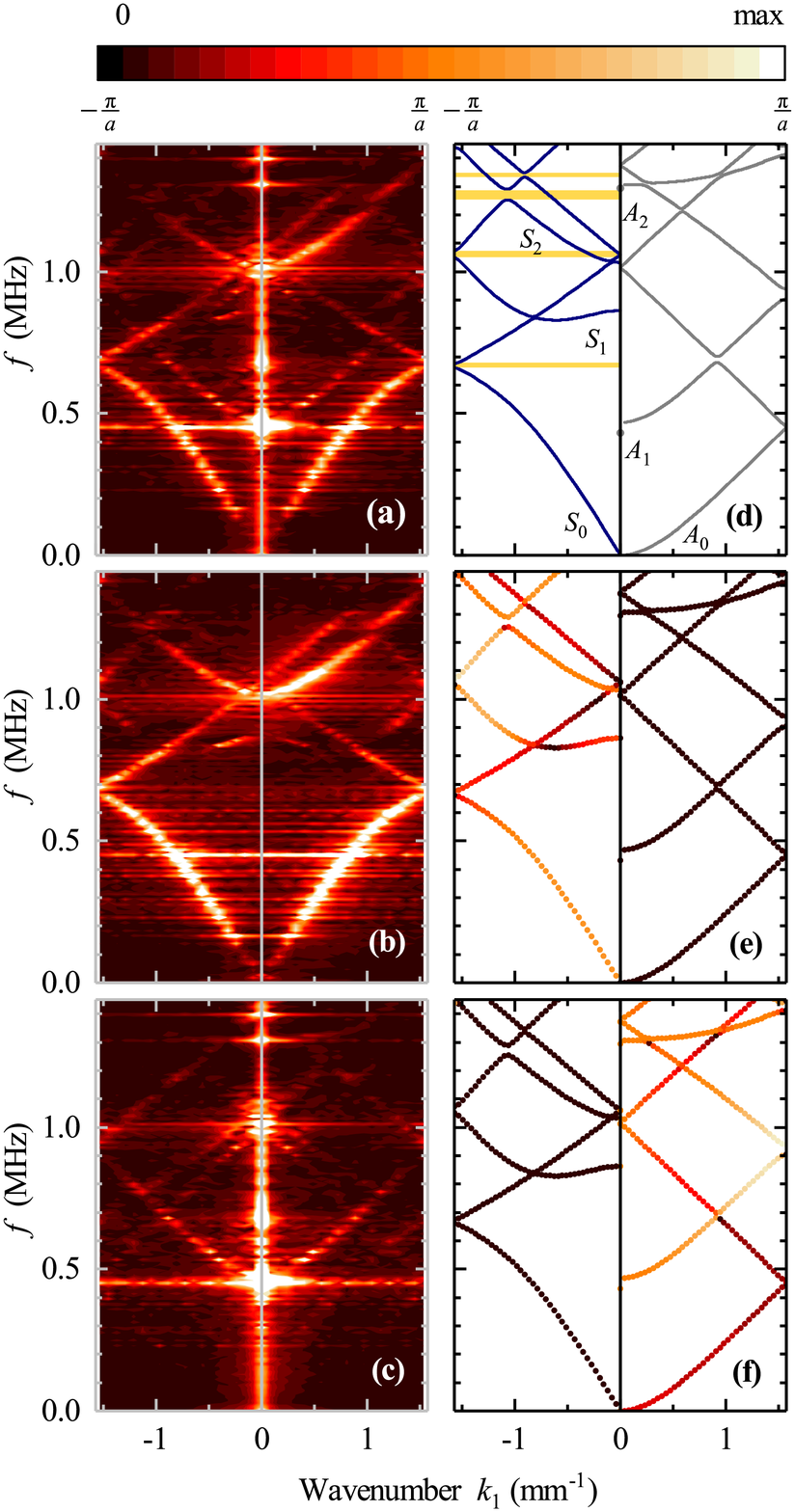}
\caption{Left panel: Experimental frequency band structure of the piezoelectric floating-potential phononic crystal plate (\emph{system I}), consisting of $N=40$ unit cells of lattice constant $a=a_0=2~\mathrm{mm}$, obtained from a double FFT of the electric-potential signal \pt(a) recorded at the upper-side electrode, $V^\mathrm{u}(x_n,t)$, and extracted from \pt(b) the difference $V^\mathrm{u}(x_n,t)-V^\mathrm{d}(x_n,t)$, and, \pt(c) the sum, $V^\mathrm{u}(x_n,t)+V^\mathrm{d}(x_n,t)$, of the electric-potential signals measured on both sides of the plate (all color maps are saturated). Right panel: The corresponding calculated band structures obtained through finite-element numerical simulations. The labels in \pt(d) correspond to usual notations adopted for Lamb modes in homogeneous plates with blue thick and gray thin lines representing, respectively, the family of symmetric and antisymmetric modes, shown separately in the negative and positive part of the $1^{\mathrm{st}}$ BZ, for a better visibility. The yellow-shaded regions indicate frequency gaps for symmetric modes in the frequency range under consideration. In \pt(e) and \pt(f) the Lamb-like modes of plot \pt(d) are colored with the difference and the sum of the calculated average potentials on the upper-side and lower-side electrodes of the unit cell.}\label{fig5}
\end{figure}
At a second stage, we show that use of the electric-potential signals provides a detailed picture of the dispersion characteristics of the Lamb-like guided modes of the finite plate (of length $l=Na_0$, $N=40$), as detailed in~\fref{fig5}; additional information on their symmetry can be also retrieved through the separated, decomposed signals, discussed in~\fref{fig4}\pt(c), \pt(d). Applying a 2D FFT on the upper-side electric-potential signal $V^\mathrm{u}(x_n,t)$, we image in~\fref{fig5}\pt(a) the frequency band structure of the crystal (a very similar picture is obtained for the lower-side potential $V^\mathrm{d}(x_n,t)$, not shown here). Comparison of this picture to the frequency band structure of the corresponding infinite crystal, calculated numerically using a finite-element commercial package~\cite{comsol} (see~\fref{fig5}\pt(d)), shows good agreement and confirms that the electric potential carries sufficient information for the imaging of the frequency band structure in such piezoelectric phononic systems.

The calculated dispersion plot of our floating-potential crystal (\emph{system I}), obtained after a careful fitting of the elastic, electric, and piezoelectric parameters (see \tref{table1}) to minimize deviation from the experimental results (\fref{fig5}\pt(a)) in the frequency range under consideration presents typical features related to the homogeneous plate dispersion behaviour, modulated by the periodicity (structured surfaces and appropriate EBCs). At the long-wavelength limit ($\omega \rightarrow 0$), in full analogy with the Lamb modes of the corresponding homogeneous plate, two branches are observed, the $S_0$-like with linear effective-medium slope ($c_{\mathrm{eff}}=3778~\mathrm{m~s^{-1}}$) and the parabolic $A_0$-like mode. Higher-frequency waveguide modes are also observed with cut-off (i.e., at $k_1=0$) frequencies: $0.47~\mathrm{MHz}$ ($A_1$-like), $0.86~\mathrm{MHz}$ ($S_1$-like), $1.03~\mathrm{MHz}$ ($S_2$-like), $1.31~\mathrm{MHz}$ ($A_2$-like). The periodic distribution of the metallic strips on both surfaces of the piezoceramic plate results, as expected, in folding of these curves at the edges and at the center of the Brillouin Zone (BZ), thus opening up narrow \emph{partial} Bragg gaps centered at $0.67$ and $1.06~\mathrm{MHz}$ ($S_0$-like), at $0.45$ and $1.01~\mathrm{MHz}$ ($A_0$-like), at $0.92$ and $1.37~\mathrm{MHz}$ ($A_1$-like), and at $1.43~\mathrm{MHz}$ ($A_2$-like). In addition, weak interactions of the Lamb-like dispersion curves (i.e., anti-crossing effects when two bands of the same symmetry meet each other) occur approximately in the middle of the BZ leading to narrow avoided-crossing (hybridization) gaps. We note in passing that there are no absolute Bragg gaps occurring in the frequency range below $1.5~\mathrm{MHz}$ for the current configuration (\emph{system I}).

\Table{\label{table1}Material parameters for PZ26, used in the calculations.}
\br
Material Parameter & Symbol & Value$^{\rm a}$\\
\mr
\multirow[c]{6}{4cm}{Elastic coefficients $c_{pq}^{E}\,[\mathrm{GPa}]$}
& $c_{11}^{E}$ & $148.0~(-11.9\%)$\\
& $c_{12}^{E}$ & $110.3~(0\%)$\\
& $c_{13}^{E}$ & $85.0~(-14.9\%)$\\
& $c_{33}^{E}$ & $135.0~(+10.1\%)$\\
& $c_{44}^{E}$ & $28.0~(-7.0\%)$\\
& ${c_{66}^{E}}^{\rm b}$ & $18.85~(-34.5\%)$\\
\mr
\multirow[c]{3}{4cm}{Piezoelectric coefficients $e_{ip}^{S}\,[\mathrm{C}\,\mathrm{m}^{-2}]$}
& $e_{15}^{S}$ & $9.86~(0\%)$\\
& $e_{31}^{S}$ & $-2.80~(0\%)$\\
& $e_{33}^{S}$ & $12.50~(-14.9\%)$\\
\mr
\multirow[c]{2}{4cm}{Relative permeability coefficients $\epsilon_{pq}^{S}$}
& $\epsilon_{11}^{S}$ & $800.0~(-3.4\%)$\\
& $\epsilon_{33}^{S}$ & $700.0~(0\%)$\\
\mr
Mass density $[\mathrm{kg}\,\mathrm{m}^{-3}]$ & $\rho$ & $7700~(0\%)$\\
\br
\end{tabular}
\item[] $^{\rm a}$ in the parenthesis, we give the relative variation with respect to the manufacturer's initial values~\cite{ferro}.
\item[] $^{\rm b}$ $c_{66}^{E}=\frac{1}{2}(c_{11}^{E}-c_{12}^{E})$.
\end{indented}
\end{table}


Moreover, the symmetric ($V^\mathrm{u}(x_n,t)-V^\mathrm{d}(x_n,t)$) and antisymmetric ($V^\mathrm{u}(x_n,t)+V^\mathrm{d}(x_n,t)$) parts of the electric potential, shown in~\fref{fig4}\pt(c), \pt(d), when processed separately, lead to two different, complementary pictures of the frequency bands corresponding to the guided modes of the periodic finite plate of symmetric and antisymmetric (with respect to the potential) character represented in~\fref{fig5}\pt(b) and \pt(c), respectively. As expected, branches that originate from symmetric Lamb modes of the homogeneous plate, such as $S_0$-, $S_1$-, and $S_2$-like ones, are manifested in the symmetric part (\fref{fig5}\pt(b)), while branches that originate from antisymmetric Lamb modes of the homogeneous plate, such as $A_0$-, $A_1$-, and $A_2$-like ones, are manifested in the antisymmetric part (\fref{fig5}\pt(c)). This picture is confirmed, in excellent agreement with the numerical calculations, if the same operation $V^\mathrm{u}\pm V^\mathrm{d}$ is applied to the potentials corresponding to each point of the dispersion plot, as witnessed by the color indexing of the dispersion lines, indicating the numerical value of difference (\fref{fig5}\pt(e)) and sum (\fref{fig5}\pt(f)) of the potential values calculated at the upper and lower strip of the unit cell. \emph{System I}, which is perfectly symmetric with respect to the $x_1x_2$-plane passing at the center of the plate (and in parallel to its surfaces), seems to support two well-defined, orthogonal subspaces of Lamb-like modes: symmetric and antisymmetric. Of course, in practise, due to finite-size effects (including off-normal excitation, i.e., small non-vanishing $k_2$-values) as well as deviations from periodicity and curvature or aberrations of the surfaces supposed to be planar and parallel in the ideal case, the corresponding modes will be an admixture of these two different classes (with a strongly, however, dominating symmetry character).

The above findings suggest that our method appears sensitive to capturing of the Lamb-like modes, in piezoelectric phononic plates, with the exception of $A_0$-like modes that seem to be less observable: this branch, as well as its folding at the center and at the edges of the BZ, is weakly projected (over ten times weaker than the rest of the modes) in the electric potential representation, in both numerical calculations and experimental results. The same trend is observed in all cases studied here (see discussion on \emph{systems II} and \emph{III}, below).

Finally, we note the presence of a strong line at $k_1=0$ in the experimental picture (see \fref{fig5}\pt(c)) extending over the whole frequency range, that originates from an instantaneously established electromagnetic excitation of all electrodes through leakage in air surrounding the structure. Along this line, some distinct points corresponding either to cut-off frequencies of some visible Lamb-like branches or to isolated solutions (at $f=0.453~\mathrm{MHz}$ for the $A_1$-like branch and at $f=1.299~\mathrm{MHz}$ for the $A_2$-like branch) are particularly over-amplified, extending, partially or entirely, along the BZ. Both these contributions are responsible for the fast waves of antisymmetric character in \fref{fig4}\pt(d).

\subsubsection{Electric-impedance analysis}

\begin{figure}[t!]
\centering
\includegraphics[width=8.2cm]{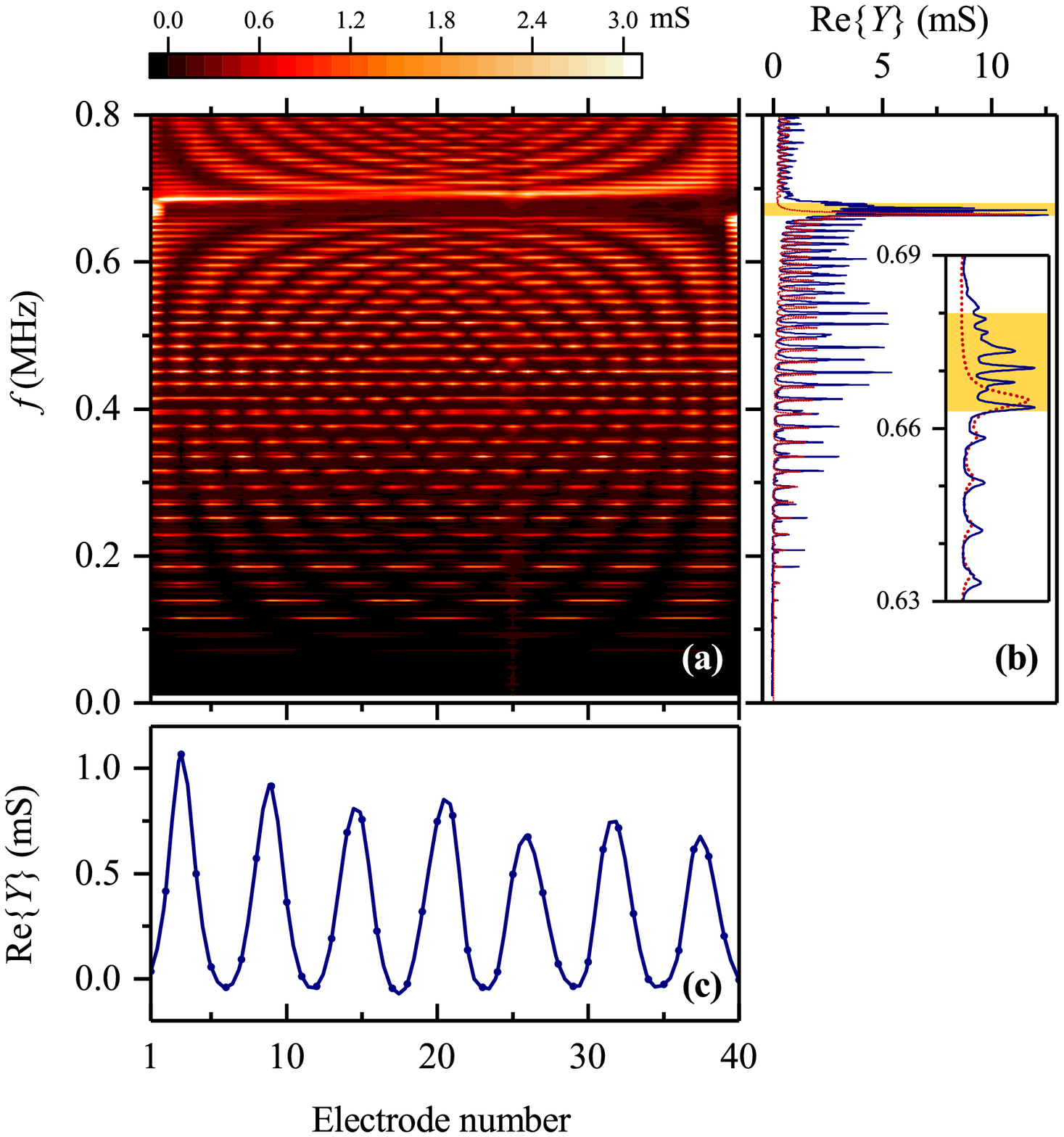}
\caption{\pt(a) Variation of the real part of the admittance measured at each pair of electrodes for the floating-potential finite crystal consisting of $N=40$ unit cells, with the position along $x_1$-direction and frequency. Two cuts of this representation \pt(b) across the $n=1$ pair of electrodes, and, \pt(c) at a given frequency $f=0.163~\mathrm{MHz}$ are also shown; blue solid and red dotted lines denote, respectively, measured and calculated values and the yellow-shaded region indicates the Bragg gap of the $S_0$-like mode.}\label{fig6}
\end{figure}

Following the procedure described in~\sref{sec:setup} we have also performed electric-impedance measurements on the sample, to complete the picture obtained from the frequency band structure, discussed previously. The impedance $Z_n$ measured at the position $x_n=(n-1)a$, i.e., at the center of the $n$-th pair of electrodes, is, as expected, sensitive to the eigenmodes of the piezoelectric phononic plate, that manifest themselves as resonance peaks in the $Z_n$ spectrum. In the absence of any external circuit loading, the floating-potential crystal (\fref{fig2}\pt(a)) can be effectively described in the low-frequency region (for frequencies below $0.30~\mathrm{MHz}$), by an equivalent planar capacitor $C_b$ whose surfaces are parallel to $x_1x_2$-plane. Indeed, the imaginary part of the admittance $Y_n=\frac{1}{Z_n}$, reveals a clear linear behavior on which several resonance structures are superimposed; for the $n=1$ (at the one edge of the plate) and $n=19$ (almost in the middle of the plate) pair of electrodes, we thus deduce, respectively, an equivalent capacitance value $C_b=0.972~\mathrm{nF}$ and $C_b=0.890~\mathrm{nF}$, while the average value on all positions gives $C_b=0.905~\mathrm{nF}$. This description is in analogy with the concept of the blocked capacitance, used to model effectively the electric behavior of homogeneous piezoelectric plates \cite{porfiri}.

\begin{figure}[t]
\centering
\includegraphics[width=7.5cm]{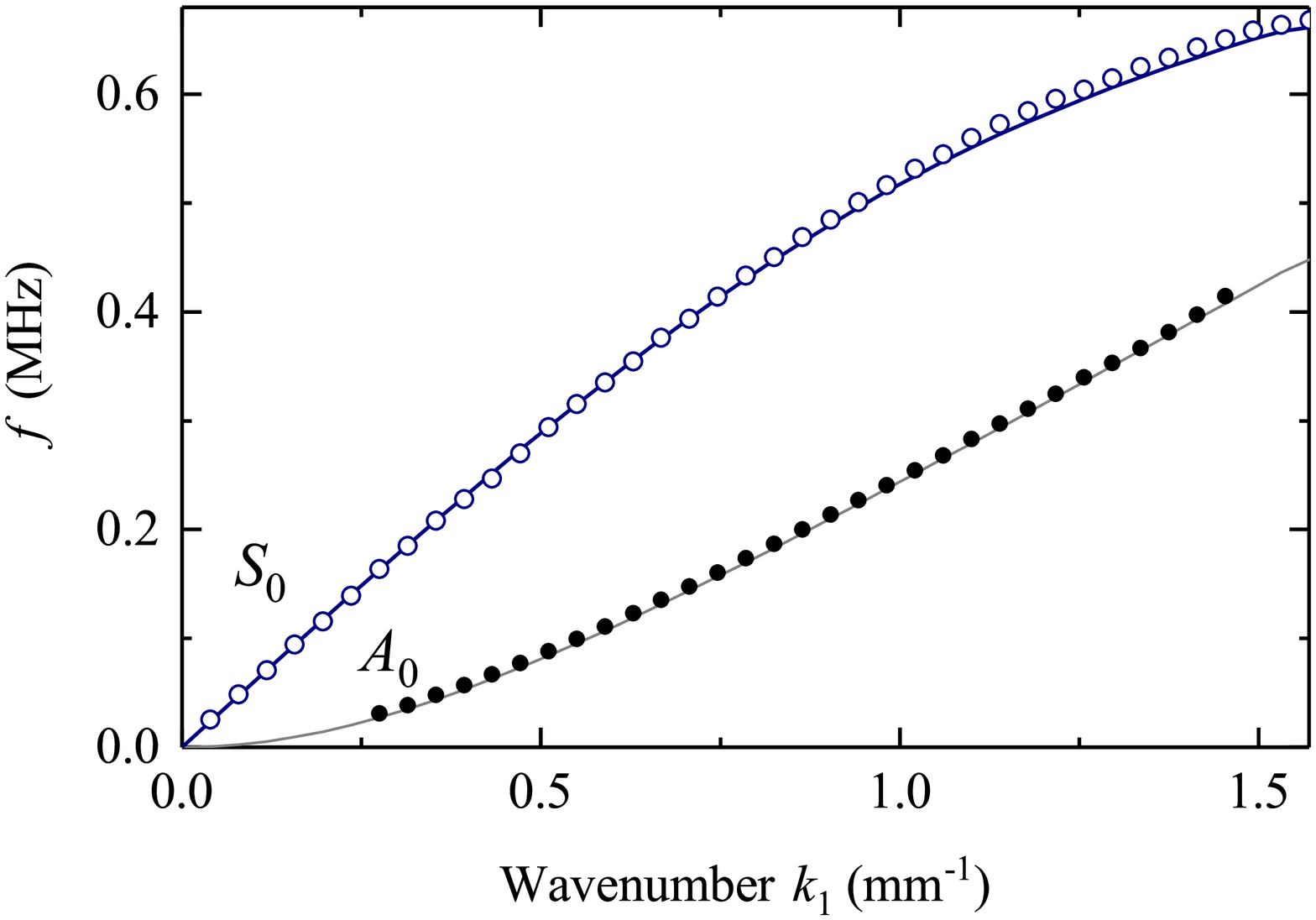}
\caption{A detail of \fref{fig5}\pt(d) showing the calculated $S_0$-like (blue line) and $A_0$-like (gray line) branches together with the experimental discrete points, retrieved, respectively, from impedance measurements along the plate thickness (open symbols) and between two adjacent electrodes (filled symbols).} \label{fig7}
\end{figure}

In accordance with the above qualitative picture, we expect $Z_n$ to be sensitive to those electromechanical modes of the crystal that correspond to significant changes in the thickness of the plate, i.e., they are symmetric with respect to the plate's median plane ($x_3=0$). We find it convenient to represent $Y_n$, rather than $Z_n$, and, more precisely, its real part, which we map onto the $(n,f)$-space in \fref{fig6}\pt(a) to better visualize the resonance positions. A closer look at the numerous sharp-resonance spectra $Y_n$ within the frequency region under consideration leads to two important conclusions: First, for frequencies below $0.47~\mathrm{MHz}$, where only two bands coexist (the $A_0$- and the $S_0$-like branches) one observes a series of high amplitude peaks centered at positions $f_m$ that can reproduce the $S_0$-like branch in a discrete manner (see open symbols in \fref{fig7}), as can be easily confirmed by applying \textquotedblleft symmetric\textquotedblright\, boundary conditions at the edges of the finite plate, along $x_1$-direction, that correspond to the quantization rule $k_{1,m}=\frac{\pi}{a}\frac{m}{N}$, where $m=1,2,\ldots,N$, $N$ being the number of strips along $x_1$, here considered $N=40$. In \fref{fig6}\pt(b) we show the measured real part of $Y_1$ (solid line) which compares well with the numerical predictions (dotted line) through finite-element simulations, where losses are taken into account by setting to $5 \cdot 10^{-3}$ the relative imaginary part of all $c_{pq}^E$ coefficients with respect to their corresponding real part. The amplitude of these peaks is modulated by a standing wave picture depending strongly on the position $x_n$ (an example is given in \fref{fig6}\pt(c) at $0.163~\mathrm{MHz}$). In addition to these peaks, a plethora of weaker-amplitude, secondary peaks occur within this frequency range that could be attributed either to the $A_0$-like eigenmodes, or to out-of-plane modes vibrating along $x_2$ axis, triggered by slight off-normal effects (such as aperture of the incident beam, possible reflections on the edges and/or corners of the sample, giving rise to secondary emissions, etc). A second remark concerns the existence of regions free of modes in the $Y_n$ spectra, whatever the position $n$ of the measurement is, thus implying that these regions are associated to frequency band gaps of the corresponding infinite phononic crystal. In the case of floating-potential crystal, we observe a region free of resonance peaks extending from $0.663$ to $0.680~\mathrm{MHz}$ that corresponds to the first Bragg gap of the $S_0$-like branch (see \fref{fig5}\pt(a)). We note in passing the presence of modes at certain frequencies which decay rapidly away from the edges of the plate, a typical example being the set of peaks lying within the frequency gap region for $S_0$-like modes (see inset in \fref{fig6}\pt(b)).

We close this part by pointing out an alternative connection of the impedance analyzer through two consecutive electrodes of the same, upper or lower, side of the plate. For frequencies below $0.45~\mathrm{MHz}$ the resonance peaks $f_m$, manifested in a $Y_n$-spectrum, reproduce well the $A_0$-like branch, always following the same quantization rule for the wavevector $k_{1,m}=\frac{\pi}{a}\frac{m}{N}$, $m=1,2,\ldots,N$, with $N=40$. The discrete points obtained in this manner for a measurement between $n$ and $n+1$ electrode positions are shown in \fref{fig7} (filled symbols). Of course, as in the case of the impedance measurement across plate's thickness, the assignment of several resonant peaks to selected, distinct bands becomes a complicated task when more than one bands coexist within a frequency window, even if these bands are not of the same symmetry (due to finite-size and off-normal effects, as we explained earlier, that favor some admixture of the modes and excitation of more than one symmetry classes at the same time).

\subsection{Symmetric EBCs including inductances connected in parallel to the plate}

\begin{figure}[t!]
\centering
\includegraphics[width=8.2cm]{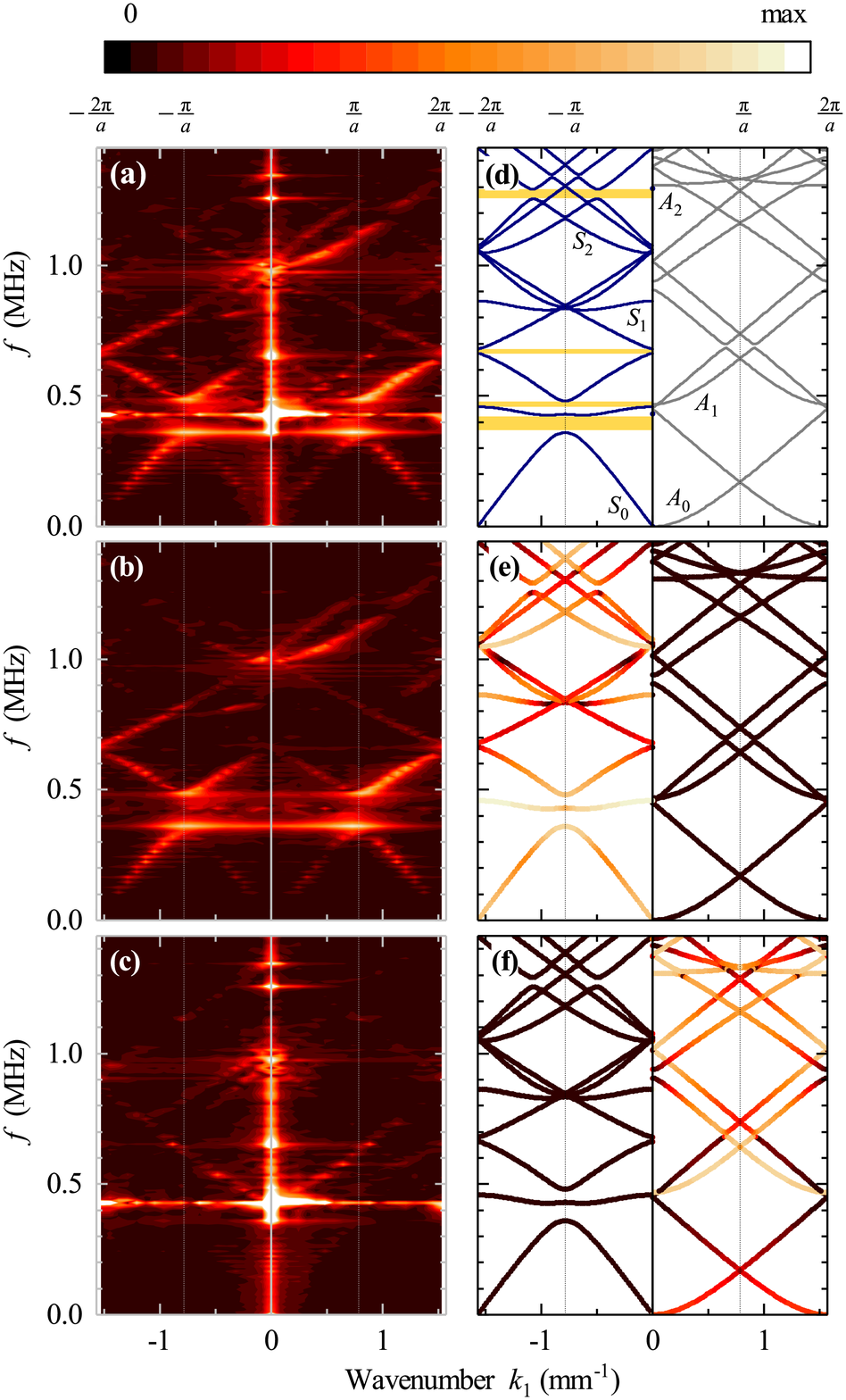}
\caption{Left panel: Experimental frequency band structure of a piezoelectric phononic crystal with EBCs depicted in \fref{fig2}\pt(b) (\emph{system II}), consisting of $N=20$ unit cells of lattice constant $a=2a_0=4~\mathrm{mm}$, obtained from a double FFT of the electric-potential signal \pt(a) recorded at the upper-side electrode, $V^\mathrm{u}(x_n,t)$, and extracted from \pt(b) the difference $V^\mathrm{u}(x_n,t)-V^\mathrm{d}(x_n,t)$, and, \pt(c) the sum, $V^\mathrm{u}(x_n,t)+V^\mathrm{d}(x_n,t)$, of the electric-potential signals measured on both sides of the plate (all color maps are saturated). Right panel: The corresponding calculated band structures obtained through finite-element numerical simulations. The labels in \pt(d) correspond to usual notations adopted for Lamb modes in homogeneous plates with blue thick and gray thin lines representing, respectively, the family of symmetric and antisymmetric modes, shown separately in the negative and positive part of the $1^{\mathrm{st}}$ and the $2^{\mathrm{nd}}$ BZ, for a better visibility. The yellow-shaded regions indicate frequency gaps for symmetric modes in the frequency range under consideration. In \pt(e) and \pt(f) the Lamb-like modes of plot \pt(d) are colored with the difference and the sum of the calculated average potentials on the upper-side and lower-side electrodes of the unit cell.} \label{fig8}
\end{figure}
Next, we introduce \emph{system II} as described in~\sref{sec:pc}, which consists of a unit cell spanning over two elementary blocks (lattice constant $a=4~\mathrm{mm}$): the EBCs are not uniform along this unit cell, depicted in \fref{fig2}\pt(b). The inductance load (we take $L=150~\mu\mathrm{H}$) is applied to even-numbered electrodes, in parallel to the plate; the structure conserves, however, the mirror symmetry with respect to the $x_1x_2$-plane (passing at the center of the plate, parallel to its two characteristic surfaces). We apply, at first, the same methodology used previously for the floating-potential crystal. The FFT of the electric-potential signals recorded at each electrode position, $x_n$, at the upper ($V^{\mathrm{u}}(x_n,t)$) and lower ($V^{\mathrm{d}}(x_n,t)$) side of the plate are practically identical and lead to a dispersion plot extending over the 1\textsuperscript{st} and 2\textsuperscript{nd} BZ (two measurement points are now included in each one of the $N=20$ unit cells constituting the finite crystal), as shown in \fref{fig8}\pt(a). The difference and sum of the upside and downside potentials, $V^{\mathrm{u}}(x_n,t)-V^{\mathrm{d}}(x_n,t)$ and $V^{\mathrm{u}}(x_n,t)+V^{\mathrm{d}}(x_n,t)$, lead respectively to two dispersion plots [see \fref{fig8}\pt(b) and \pt(c)] decomposing the band structure of \fref{fig8}\pt(a) into a symmetric and an antisymmetric part. To facilitate the analysis of these plots we compute the frequency band structure of the corresponding infinite crystal, shown, respectively, in \fref{fig8}\pt(d), \pt(e), and \pt(f). One expects that doubling of the unit-cell length, as compared to those of \emph{system I}, will induce folding of the bands of the floating-potential crystal at $k_1=\frac{\pi}{2a_0}=785.4~\mathrm{m}^{-1}$. Moreover, the inductance load will add new, localized modes originating from $LC$-like electric resonances which are generated by the coupling of the inductance $L$, loaded at each cell, with a virtual capacitor (or a combination of capacitors) describing effectively the electromagnetic coupling that takes place within the piezoceramic plate. Indeed, the above picture is confirmed in \fref{fig8}\pt(d), and the flat band centered at about $0.45~\mathrm{MHz}$ seems to be a good candidate for localized modes of electric-resonator origin.

Using the same operation as in the right panel of \fref{fig5} (plots \pt(e) and \pt(f)), we observe that the difference (\fref{fig8}\pt(e)) and the sum (\fref{fig8}\pt(f)) are non-vanishing over two separate classes of modes of the phononic crystal piezoelectric plate, that coincide with the symmetric- and antisymmetric-character labeling of the frequency bands of the corresponding homogeneous plate. The configuration under study (\emph{system II}) inherits this feature from the initial (homogeneous plate) system since the horizontal mirror symmetry is not destroyed. Good agreement is obtained between the experimental results and the numerical predictions, showing that the flat band is of symmetric character (see \fref{fig8}\pt(e)). In \fref{fig8}\pt(b) the $S_0$-, $S_1$-, and $S_2$-like modes are clearly identified. More importantly, two flat modes, at frequencies $0.374$ and $0.503~\mathrm{MHz}$, are observed in the experimental picture, not visible in the numerical dispersion plot of \fref{fig8}\pt(e) (they will be discussed below). Concerning the antisymmetric subspace, as in the case of \emph{system I} the $A_0$-like branch is discernible only after its first folding and close to the center of the 1\textsuperscript{st} BZ, at about $1.00~\mathrm{MHz}$, while the $A_1$-like branch is easily identified.

\begin{figure}[t]
\centering
\includegraphics[width=8.2cm]{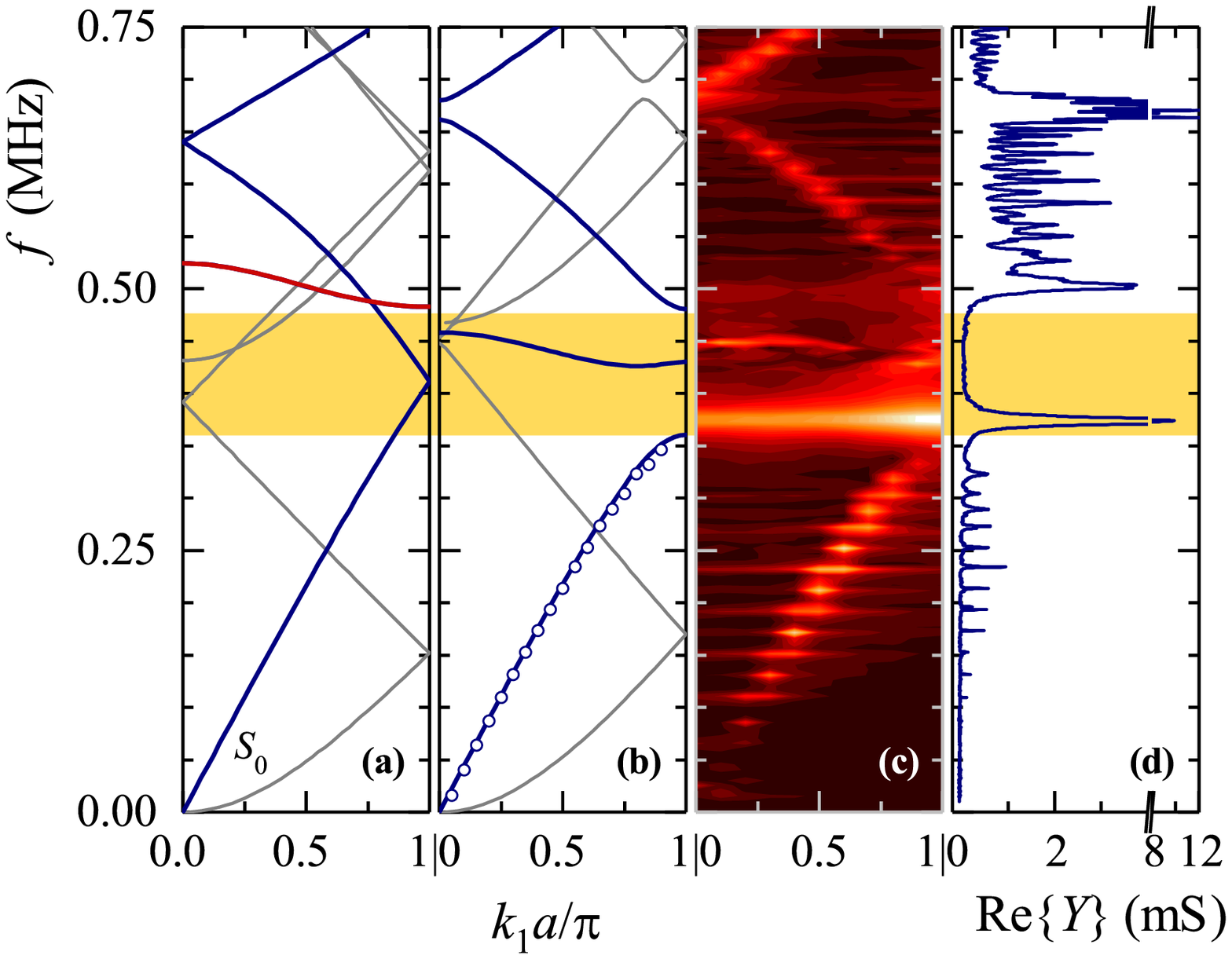}
\caption{Calculated band structure of the crystal shown in \fref{fig2}\pt(b) (\emph{system II}) with \pt(a) no electromechanical coupling and \pt(b) full electromechanical coupling taken into account. Color lines follow the same nomenclature as in \fref{fig8} and the red line shows the numerically fitted electric band of the corresponding transmission-line model given by \eref{band1}. The experimental frequency band structure obtained from a 2D FFT of the difference of the up- and down-side potential signals measured at the even-pair electrodes only is shown in \pt(c). The real part of the admittance $Y$, shown in \pt(d), is measured at $n=1$, whose peak positions reproduce the $S_0$-like band in \pt(b) (open symbols). The yellow-shaded region denotes the calculated band gap region of plot \pt(b).} \label{fig9}
\end{figure}
We shall further analyze the symmetric part of the dispersion plot for frequencies below $0.75~\mathrm{MHz}$. In \fref{fig9}\pt(a) we give a detail of \fref{fig8}\pt(d) and in \fref{fig9}\pt(b) we compute the frequency band structure for the crystal under study, after having switched off any electromechanical coupling: all $e_{ip}^{S}$ coefficients are set to zero and one expects that any electric-resonant band resulting from the electric interaction of the $LC$-like resonators will appear in the dispersion diagram in its unhybridized form, i.e., without any interaction occurring with the Lamb-like elastic modes of the phononic plate. After a careful comparison of \fref{fig9}\pt(a) with the folded at $k_1=785.4~\mathrm{m}^{-1}$ corresponding dispersion plot of \emph{system I} we conclude that the flat band extending from $0.482~\mathrm{MHz}$ ($\frac{k_1a}{\pi}=1$) to $0.524~\mathrm{MHz}$ ($\frac{k_1a}{\pi}=0$) is the electric-resonance band; the rest of the frequency bands correspond to typical Lamb-like modes. In the real system (see \fref{fig9}\pt(b)), these modes lower slightly in frequency, due to the modification in the electromechanical parameters entering in the eigenvalue problem (see~\cite{kherraz3}) after switching off the coupling terms. To confirm the electric origin of this flat mode, an equivalent electric-line model is developed, to describe and reproduce the coupling of the isolated electric resonators, arranged in a periodic chain (see Appendix). The unit cell is described by a coupling capacitor $C$ connecting the upper and lower lines, each one corresponding to the upper and lower series of structured on the surfaces of the plate metallic strips, which are described by planar capacitors $C_{\mathrm{s}}$ oriented in parallel to $x_1x_3$-plane; the equivalent-circuit unit cell is depicted in \fref{figA1}\pt(a). After a straightforward application of Kirchhoff's current and voltage laws together with Bloch's theorem we find a cos-like dispersion line $\omega(k_1)$ describing this narrow electric-resonant band, given by
\begin{equation}\label{band1}
\omega=\omega_0\left(1+\frac{C_{\mathrm{s}}}{C}\sin^2\frac{k_1a}{2}\right)^{-\frac{1}{2}}\,,
\end{equation}
where $\omega_0=\frac{1}{\sqrt{LC}}$ is the angular frequency at the center of the BZ ($k_1=0$). The equivalent $C$ and $C_{\mathrm{s}}$ parameters are easily determined through a non-linear point-to-point fitting of the computed resonant band, shown in red in~\fref{fig9}\pt(a), with~\eref{band1}. We obtain $C=0.614~\mathrm{nF}$ and $C_{\mathrm{s}}=0.110~\mathrm{nF}$, valid along the whole electric band.

In the real system the electromechanical coupling must be taken into account and the flat electric band will interact with Lamb-like modes of the same symmetry, leading to hybridization band gaps (HBG) opening up at these crossing regions. In our case, the electric-resonant band, as a result of the coupling through non-vanishing $e_{ip}^{S}$ coefficients, shifts at lower frequencies and interacts with the first $S_0$-like branch and its folding. Accidentally, the HBG occurs at the vicinity of the $S_0$ first Bragg gap, thus leading to a significant widening of the gap region that now extends from $0.360$ to $0.481~\mathrm{MHz}$ interrupted by the flat band, lying in its interior from $0.426$ to $0.458~\mathrm{MHz}$. The presence of this \emph{partial} (for symmetric modes only) wide gap is confirmed in the experimental dispersion plot~(\fref{fig8}\pt(b)), which is not the case for the flat electric-resonant band, centered at $0.503~\mathrm{MHz}$ in the computed dispersion plot of~\fref{fig9}\pt(b), hardly seen in the experimental band structure. It could be possibly shadowed by other high-amplitude contributions. We remind that $L$ loads are connected to the even-pair electrodes, in our crystal; for this reason, we repeat the FFT of~\fref{fig8}\pt(b) by taking into account only the even-numbered electrode positions. The result is shown in~\fref{fig9}\pt(c), revealing clearly the presence of the gap (from $0.373$ to $0.500~\mathrm{MHz}$) together with a part of the flat resonant band lying within this frequency region, extending from $0.433$ to $0.454~\mathrm{MHz}$. It worths remembering that the flat band is an admixture of $S_0$-like and electric-resonance modes, the points lying at the first $\frac{3}{4}$ along this line have stronger electric-resonance character. For those points we know (and this has been confirmed in our numerical simulations) that the deformation along the thickness direction ($x_3$) of the plate is much higher at the part of the unit cell where $L$ is loaded (right half) and much weaker at the other part (left half) of the unit cell where floating-potential EBCs are applied (see \fref{fig2}\pt(b)).

\begin{figure}[t!]
\centering
\includegraphics[width=8cm]{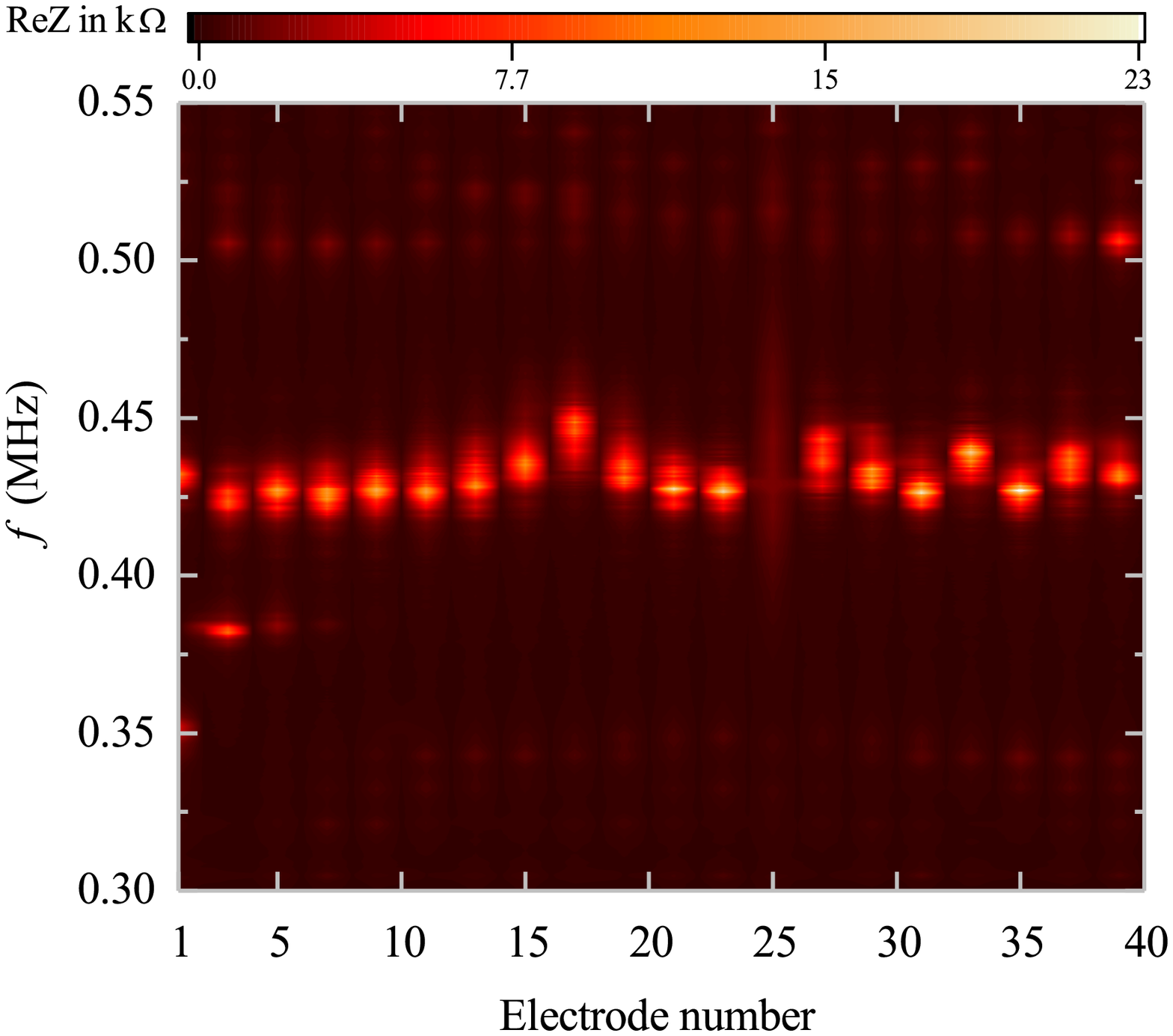}
\caption{Evolution of the real part of the electric impedance with electrode position, measured along $x_1$ direction of \emph{system II}, within the gap region shown in \fref{fig9}.}\label{fig10}
\end{figure}
We corroborate this picture with the help of the electric impedance measured at the first electrode position. Precisely, the real part of the admittance shown in~\fref{fig9}\pt(d) reveals a number of sharp, narrow peaks for frequencies below $0.346~\mathrm{MHz}$, whose assignment to discrete wavenumber values following the rule $k_{1,m}=\frac{\pi}{a}\frac{m}{N}$, $m=1,2,\ldots,N$, where $N=20$, reproduces well the $S_0$-like branch (see symbols in \fref{fig9}\pt(b)). A region free of resonance peaks extending from $0.392$ to $0.484~\mathrm{MHz}$ corresponds to the gap, predicted by the numerical simulations, which is delimited by two strong peaks in the $Y$ spectrum at $0.374$ and $0.503~\mathrm{MHz}$. The inverse electric-potential FFT of each one of these two selected flat bands shows that they correspond to modes of wavelength $\lambda\approx4a_0$, localized in the first few unit cells of the crystal, close to the excitation point. The spectra of admittance $Y_n$ measured along $x_1$ direction at positions $x_{n}$ do not allow an identification of the flat electric band, perturbed by several contributions related to interference phenomena, sensitive to the presence of the inductance $L$. The impedance seems to provide a more clear picture, in accordance with the results found from the electric-potential analysis. In \fref{fig10} we plot the evolution of the real part of the impedance, recorded along $x_1$ direction, as a function of frequency, within the gap region of \fref{fig9}\pt(b). One observes the presence of several high-amplitude resonance structures, each containing more than one resonance peaks that cannot be separately resolved. They form a resonant band extending from $0.418$ to $0.455~\mathrm{MHz}$, which corresponds well to the frequency region found in \fref{fig9}\pt(c).

\subsection{Non-symmetric EBCs including connected inductances in series to the plate}

As a last case, we examine \emph{system III} described in~\sref{sec:pc}, which, again, consists of a unit cell spanning over two elementary blocks (lattice constant $a=4~\mathrm{mm}$), with the difference that now the EBCs, depicted in \fref{fig2}\pt(c), are not symmetric with respect to the median horizontal plane of the plate: floating-potential conditions are applied to the left part (odd-numbered electrodes) of the unit cell (as in \emph{system II}); the right part (even-numbered electrodes) is loaded with an inductance $L=150~\mu\mathrm{H}$ connected in series with the plate at the lower-side electrode, while floating-potential conditions are applied at the upper-side electrode.

\begin{figure}[t!]
\centering
\includegraphics[width=7.3cm]{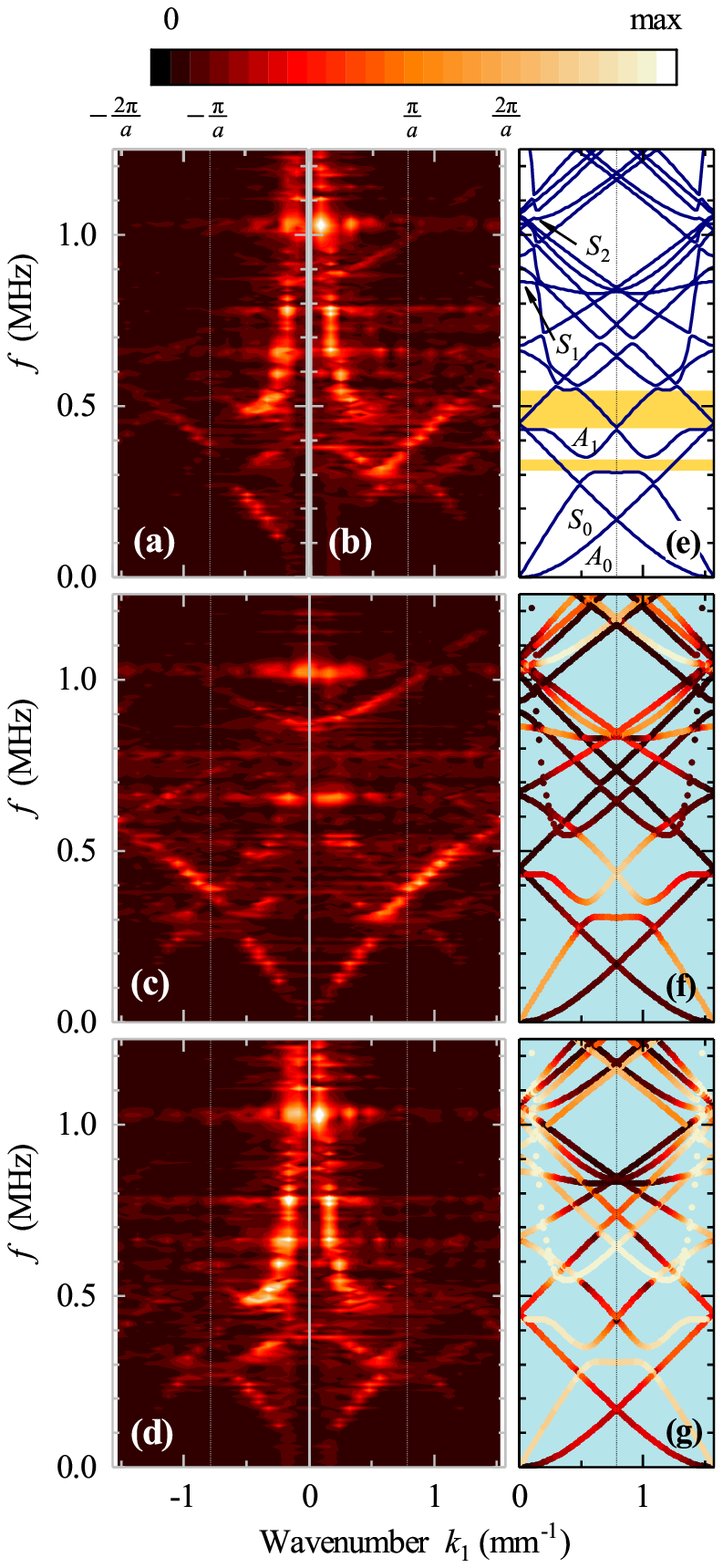}
\caption{Left panel: Experimental frequency band structure of a piezoelectric phononic crystal with EBCs depicted in \fref{fig2}\pt(c) (\emph{system III}), consisting of $N=20$ unit cells of lattice constant $a=2a_0=4~\mathrm{mm}$, obtained from a double FFT of the electric-potential signal recorded at the \pt(a) upper-side and \pt(b) lower-side electrode, $V^\mathrm{u}(x_n,t)$ and $V^\mathrm{d}(x_n,t)$, respectively, and extracted from \pt(c) the difference $V^\mathrm{u}(x_n,t)-V^\mathrm{d}(x_n,t)$, and, \pt(d) the sum, $V^\mathrm{u}(x_n,t)+V^\mathrm{d}(x_n,t)$, of the electric-potential signals measured on both sides of the plate (all color maps are saturated). Right panel: The corresponding calculated band structures obtained through finite-element numerical simulations. The labels in \pt(e) correspond to usual notations adopted for Lamb modes in homogeneous plates. The yellow-shaded regions denote avoided crossings that originate from the interaction of the electric resonant modes with the $S_0$-like and $A_1$-like guided modes of the plate. In \pt(f) and \pt(g) the Lamb-like modes of plot \pt(e) are colored with the difference and the sum of the calculated average potentials on the upper-side and lower-side electrodes of the unit cell.} \label{fig11}
\end{figure}
We repeat the same steps in our analysis, concerning the dispersion plot of this system. Again, we retrieve the experimental frequency band structure after a double FFT is applied to the electric-potential signals recorded at each electrode position of the upper ($V^{\mathrm{u}}(x_n,t)$) or the lower ($V^{\mathrm{d}}(x_n,t)$) side of the plate: the corresponding pictures, shown respectively in \fref{fig11}\pt(a) and \pt(b), differ, as one could expect, lack of mirror symmetry with respect to $x_1x_2$-plane. The main differences lie within the frequency range below $0.45~\mathrm{MHz}$ and concern the weakening of the $S_0$-like branch for the case of lower-side measurements, below the first HBG extending from $0.312$ to $0.343~\mathrm{MHz}$, while the band just above this gap extending from $0.343$ to $0.442~\mathrm{MHz}$ and an almost flat part of a dispersion line at $f\simeq0.300~\mathrm{MHz}$ seem to be enhanced. The above observations suggest that these modes are strongly related to the presence of the inductance load and should contain an important percentage of electric-resonance hybridization. In \fref{fig11}\pt(c) and \pt(d) we show, as usually, the dispersion plot obtained from the FFT of the difference, $V^{\mathrm{u}}(x_n,t)-V^{\mathrm{d}}(x_n,t)$, and the sum, $V^{\mathrm{u}}(x_n,t)+V^{\mathrm{d}}(x_n,t)$, of the separately recorded signals at the two sides of the plate. The corresponding computed frequency band structures, with elastic, electric and piezoelectric constants taken again from~\tref{table1}, are shown in \fref{fig11}\pt(e), \pt(f) and \pt(g), using the same representation: we map along these curves the computed difference and sum of the electric-potential values, on the upper- and lower-side electrodes of each unit cell; the results are in good agreement with the experimental picture.

The most striking difference, as compared to the two previously studied systems is the absence of crossing between any two dispersion lines: every time two curves cross each other an avoided crossing takes place and an interaction occurs (whatever the degree of this interaction: strong or weak). This implies that all bands must belong to the same symmetry class, i.e., their classification into symmetric and antisymmetric modes is not valid any more, since \emph{system III} lowers its symmetry because of the one-side inductance load. This is also confirmed by the picture given in \fref{fig11}\pt(f) and \pt(g). Each eigenmode contains both \textquotedblleft symmetric\textquotedblright ($V^{\mathrm{u}}-V^{\mathrm{d}}$) and \textquotedblleft antisymmetric\textquotedblright ($V^{\mathrm{u}}+V^{\mathrm{d}}$) parts of the potentials and is ---in a bigger or lesser degree--- observed in both representations. As an example, the $S_0$-like branch is not anymore purely symmetric, but an admixture of symmetric and antisymmetric contributions. That is why, strictly speaking, labeling of the several dispersion lines with terms used for the modes in homogeneous plates is not appropriate, in this case, and should be avoided. Finally, the results presented in \fref{fig11} reveal the existence of an unusual curve with an almost vertical part at $k_1\simeq 0.125~\mathrm{mm}^{-1}$ for frequencies higher than $0.55~\mathrm{MHz}$. This branch carries a strong antisymmetric hybridized character and could be of electric-resonant origin, induced by the presence of the $L$-loads that couple with the piezoelectric plate which is effectively described by an equivalent system of planar capacitors.

\begin{figure}[t]
\centering
\includegraphics[width=7.5cm]{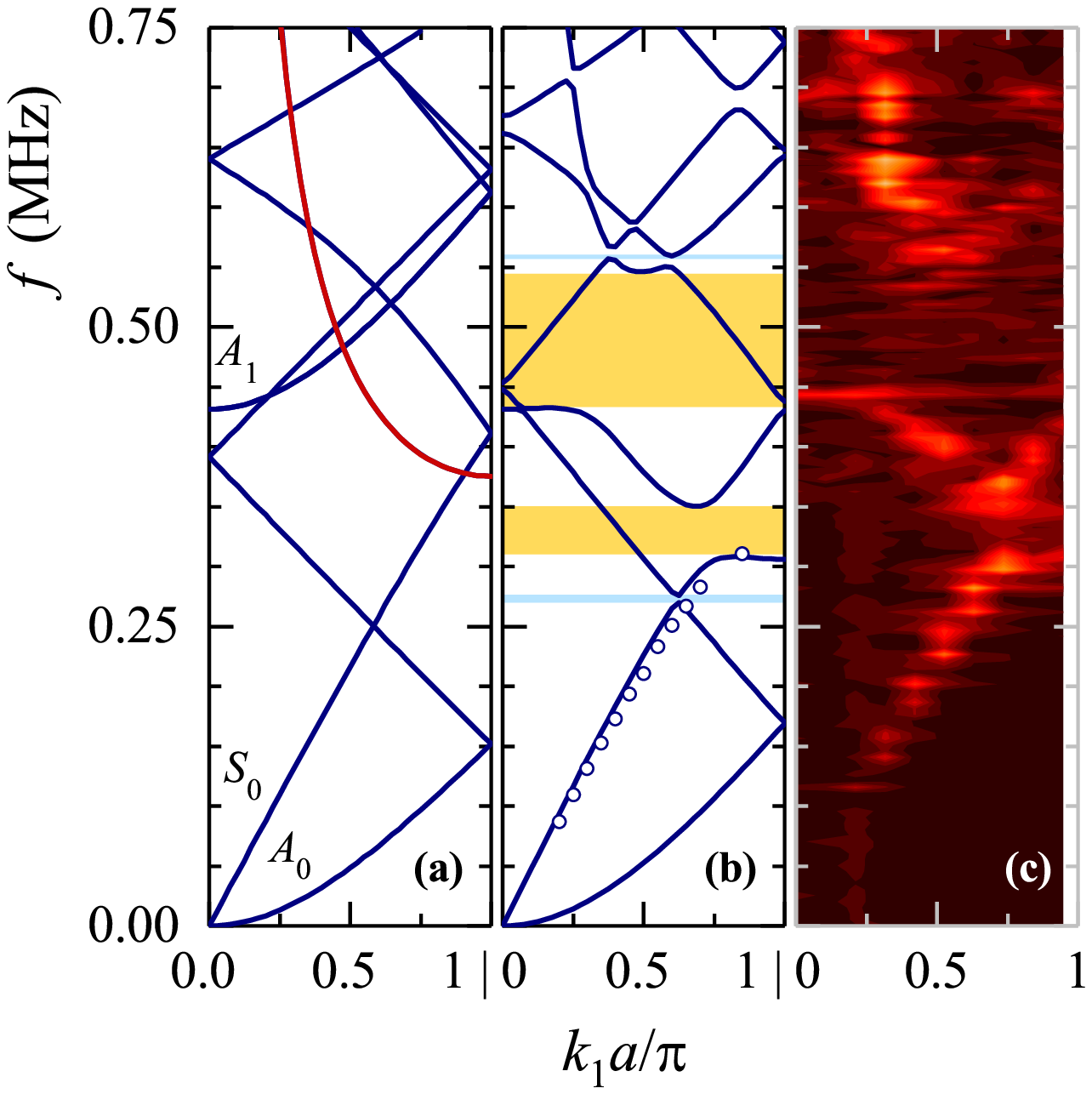}
\caption{Calculated band structure of the crystal shown in \fref{fig2}\pt(c) (\emph{system III}) with \pt(a) no electromechanical coupling and \pt(b) full electromechanical coupling taken into account. Color lines follow the same nomenclature as in \fref{fig11} and the red line shows the numerically fitted electric band of the corresponding transmission-line model given by \eref{band2}. In \pt(c) the experimental frequency band structure obtained from a 2D FFT of the sum of the up- and down-side potential signals measured at the odd-pair electrodes only is given for comparison. The resonant peaks of the real part of the admittance $Y_1$ are used to reproduce the $S_0$-like band in \pt(b) (open symbols). The yellow-shaded and blue-shaded regions denote, respectively, avoided crossings (see text) and absolute frequency band gaps.} \label{fig12}
\end{figure}
To clarify this point, we calculate the frequency band structure of \emph{system III} when the electromechanical coupling is switched off (by setting $e_{ip}^{S}=0$). The results, shown in \fref{fig12}\pt(a) reveal, indeed, a very wide, hyperbola-like band with zero group velocity at $0.374~\mathrm{MHz}$ (at the edge of the $1^\mathrm{st}$ BZ), that originates from the coupling of the individual $LC$-like resonators; the remaining elastic branches coincide to those of \emph{system II}, depicted in \fref{fig9}\pt(a). An attempt to describe this electric band may be realized through an equivalent transmission-line periodic model whose unit cell is given in \fref{figA1}\pt(b) (see Appendix). After a straightforward application of Kirchhoff's current and voltage laws together with Bloch's theorem we find a hyperbola-like dispersion line $\omega(k_1)$ describing this very wide electric-resonant band, given by
\begin{eqnarray}\label{band2}
\omega&=&\omega_0\left[1+2\frac{C_{\mathrm{s}}^{\mathrm{d}}}{C}\sin^2\frac{k_1a}{2}\right.\nonumber \\ &-&
\left.\left(1+2\frac{C_{\mathrm{s}}^{\mathrm{u}}}{C}\sin^2\frac{k_1a}{2}\right)^{-1}\right]^{-\frac{1}{2}}\,,
\end{eqnarray}
where again $\omega_0=\frac{1}{\sqrt{LC}}$, while for wavevector values close to the center of the BZ the angular frequency is approximately given by $\omega\left(\frac{k_1a}{\pi}\ll1\right)\approx\frac{\omega_1}{k_1a}$, with $\omega_1=\frac{1}{\sqrt{L\frac{(C_{\mathrm{s}}^{\mathrm{u}}+C_{\mathrm{s}}^{\mathrm{d}})}{2}}}$. The equivalent $C$, $C_{\mathrm{s}}^{\mathrm{u}}$ and $C_{\mathrm{s}}^{\mathrm{d}}$ parameters are easily determined through a non-linear point-to-point fitting of the computed resonant band, shown in red in \fref{fig12}\pt(a), with~\eref{band2}. We obtain $C=1.658~\mathrm{nF}$, $C_{\mathrm{s}}^{\mathrm{u}}=0.983~\mathrm{nF}$ and $C_{\mathrm{s}}^{\mathrm{d}}=0.149~\mathrm{nF}$, valid along the whole electric band.

Next, we switch on the electromechanical coupling (see \fref{fig12}\pt(b), which corresponds to a detail of \fref{fig11}\pt(e)). The very wide electric-resonance band couples with the Lamb-like modes of the phononic plate and interacts strongly with the $A_1$- and in a lesser degree with the $S_0$-like mode of the plate, thus leading to large avoided crossings that extend, respectively, from $0.308$ to $0.350~\mathrm{MHz}$, and from $0.432$ to $0.546~\mathrm{MHz}$. Finally, the appearance of very narrow, absolute HBGs is also observed (see blue-shaded regions in \fref{fig12}\pt(b)), centered at $0.275$ and $0.558~\mathrm{MHz}$, as a result of the weak interaction between the $S_0$ and $A_0$-like modes, the former, and between the unhybridized electric-resonance band (red dotted lines) and some guided modes of the plate, the latter.

It is worth noting that, though $L$-circuit loads have been already used in the past~\cite{kherraz2}, the configuration under study provides unprecedented characteristics in the dispersion relation plots, related to the very wide electric resonant band, whose low-symmetry character (a hybridization of dominating antisymmetric with symmetric modes) leads to a strong interaction with the $A_1$-like mode and in a lesser degree with $S_0$; the inverse was observed for \emph{system II}.

\section{Conclusion}~\label{sec:conclusion}

In summary, we developed an experimental methodology based on all-electric measurements to retrieve and analyze, in terms of symmetry, the frequency band structure of a piezoelectric phononic plate structured on the surfaces with periodically distributed arrays of metallic strips, used for both excitation and reception of the response of the plate. This method, examined in three model cases including symmetric and non-symmetric applied EBCs with several inductance-load configurations, has been shown to be robust, efficient, and, reliable in providing a rich information on the symmetry of the electromechanical modes and, at the same time, promising in terms of invariant performances under rescaling, especially for structure miniaturization. More importantly, our results, in good agreement with numerical predictions, demonstrate the possibility of inducing new, unusual electric modes that can interact at will with the guided Lamb-like modes generated in the plate, through a variety of combinations that the external electric circuits may offer. Lowering in a non-destructive manner the symmetry of the structure, by means of external loads, controls the symmetry of the dispersion curves, thus leading to possible openings of avoided-crossing gaps, whose width and degree of interaction should be studied in the future.

\ack
F.-H.C.-B. was supported by the Normandy Region RIN program METACAP through a postgraduate fellowship.

\appendix
\setcounter{section}{1}
\section*{Appendix: Equivalent electric circuits}

\begin{figure}[t]
\centering
\includegraphics[width=6.5cm]{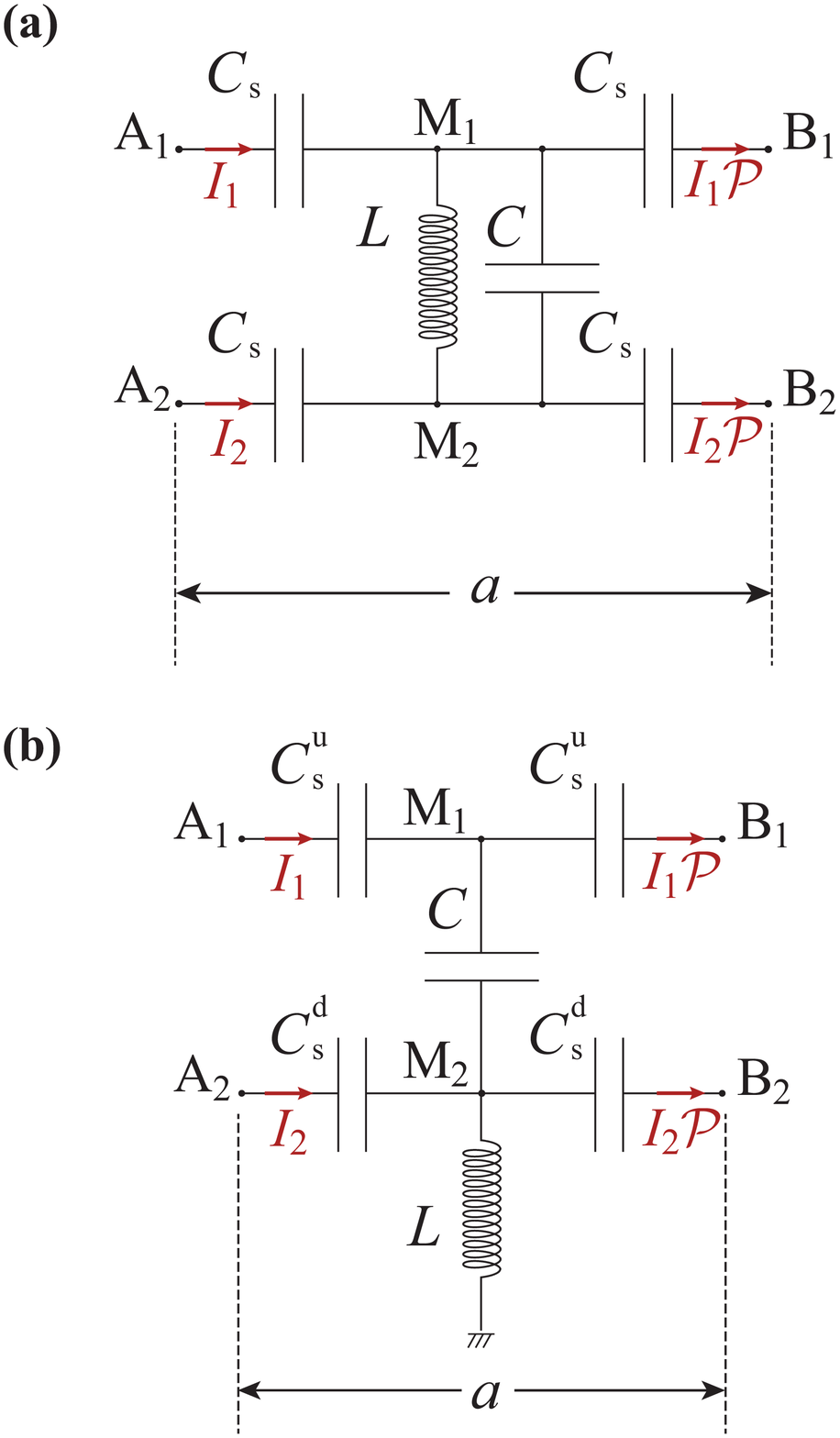}
\caption{The unit cell of a periodic transmission line describing equivalently \pt(a) \emph{system II} (\fref{fig2}\pt(b)) and \pt(b) \emph{system III} (\fref{fig2}\pt(c)) when all electromechanical couplings are switched off.}\label{figA1}
\end{figure}

The equivalent circuit that models the electric resonator of \emph{system II} when the piezoelectric coupling is switched off is depicted in \fref{figA1}\pt(a). We define all voltages in the input, $V_{A_j}$, and output, $V_{B_j}=\mathcal{P}V_{A_j}$, of the unit cell, as well as those at the edges of the inductance load, $V_{M_j}$, with respect to a common ground reference, where $j=1,2$, and $\mathcal{P}=\mathrm{e}^{-\mathrm{i}k_1a}$ is the Bloch phase factor (we assume an $\mathrm{e}^{+\mathrm{i}\omega t}$ time dependance in all fields); the incoming and outgoing electric currents are respectively $I_j$ and $\mathcal{P}I_j$. Application of Kirchhoff's current and voltage laws gives
\begin{eqnarray}\label{kirch1}
  \frac{V_{A_j}-V_{M_j}}{I_j} &=& \frac{V_{M_j}-V_{A_j}\mathcal{P}}{I_j\mathcal{P}}=Z_{\mathrm{s}},\quad j=1,2, \nonumber \\
  \frac{V_{M_1}-V_{M_2}}{I_1(1-\mathcal{P})}&=& \frac{V_{M_2}-V_{M_1}}{I_2(1-\mathcal{P})}=\frac{Z_CZ_L}{Z_C+Z_L},
\end{eqnarray}
where we defined $Z_{\mathrm{s}}=\frac{1}{\mathrm{i}C_{\mathrm{s}}\omega}$, $Z_C=\frac{1}{\mathrm{i}C\omega}$, and $Z_L=\mathrm{i}L\omega$. From~\eref{kirch1} one can easily see that $I_1=-I_2$, $V_{A_1}=-V_{A_2}$, $V_{M_1}=-V_{M_2}$, and $V_{M_j}=V_{A_j}\frac{2\mathcal{P}}{1+\mathcal{P}}$, that finally reduce the above system to a $2\times2$ form which, after elimination of the ratio $\frac{V_{A_1}}{I_1}$, leads to a secular equation depending only on $\mathcal{P}$ and the set of unit-cell impedances. We thus obtain the following dispersion equation for the electric resonant band
\begin{equation*}
\omega=\omega_0\left(1+\frac{C_{\mathrm{s}}}{C}\sin^2\frac{k_1a}{2}\right)^{-\frac{1}{2}}\,,
\end{equation*}
where $\omega_0=\frac{1}{\sqrt{LC}}$ is the angular frequency at the center of the BZ ($k_1=0$); at the edge of the BZ ($k_1=\frac{\pi}{a}$) we find $\omega_{\pi}=\frac{1}{\sqrt{L(C+C_{\mathrm{s}})}}$. The width $\Delta\omega=\omega_0-\omega_{\pi}$ of this cos-like resonant band is finite.

To model the unit cell of the asymmetric with respect to $x_1x_2$-plane crystal, we will assume the capacitors $C_{\mathrm{s}}$ to be different for the upper and lower side (see \fref{figA1}\pt(b)). We follow the same considerations for the quantities concerning the left and right ports, by applying the Bloch phase factor, and application of Kirchhoff's current and voltage laws gives
\begin{eqnarray}\label{kirch2}
  \frac{V_{A_j}-V_{M_j}}{I_j} &=& \frac{V_{M_j}-V_{A_j}\mathcal{P}}{I_j\mathcal{P}}=Z_{\mathrm{s}}^{\nu}, \nonumber \\
  \frac{V_{M_1}-V_{M_2}}{I_1(1-\mathcal{P})}&=& Z_C, \nonumber \\
  \frac{V_{M_2}}{(I_1+I_2)(1-\mathcal{P})}&=& Z_L.
\end{eqnarray}
Here we defined $Z_{\mathrm{s}}^{\nu}=\frac{1}{\mathrm{i}C_{\mathrm{s}}^{\nu}\omega}$ with $\nu=\mathrm{u}(\mathrm{d})$ for $j=1(2)$, while $Z_C=\frac{1}{\mathrm{i}C\omega}$ and $Z_L=\mathrm{i}L\omega$, as in the previous case. From~\eref{kirch2} one can easily see that $I_j=V_{A_j}\frac{1}{Z_{\mathrm{s}}^{\nu}}\frac{1-\mathcal{P}}{1+\mathcal{P}}$, and $V_{M_j}=V_{A_j}\frac{2\mathcal{P}}{1+\mathcal{P}}$, that finally reduce the above system to a $2\times2$ form which, after elimination of the unknowns $V_{A_j}$, leads to a secular equation depending only on $\mathcal{P}$ and the set of unit-cell impedances. We thus obtain the following dispersion equation for the electric resonant band
\begin{eqnarray*}
\omega&=&\omega_0\left[1+2\frac{C_{\mathrm{s}}^{\mathrm{d}}}{C}\sin^2\frac{k_1a}{2}\right.\nonumber \\ &-&
\left.\left(1+2\frac{C_{\mathrm{s}}^{\mathrm{u}}}{C}\sin^2\frac{k_1a}{2}\right)^{-1}\right]^{-\frac{1}{2}}\,,
\end{eqnarray*}
where again $\omega_0=\frac{1}{\sqrt{LC}}$. For wavevector values close to the center of the BZ the angular frequency is approximately given by $\omega\left(\frac{k_1a}{\pi}\ll1\right)\approx\frac{\omega_1}{k_1a}$, with $\omega_1=\frac{1}{\sqrt{L\frac{(C_{\mathrm{s}}^{\mathrm{u}}+C_{\mathrm{s}}^{\mathrm{d}})}{2}}}$, diverging according to an hyperbola-like behaviour; at the edge of the BZ ($k_1=\frac{\pi}{a}$) we find $\omega_{\pi}=\omega_0\left[1+2\frac{C_{\mathrm{s}}^{\mathrm{d}}}{C}-\left(1+2\frac{C_{\mathrm{s}}^{\mathrm{u}}}{C}\right)^{-1}\right]^{-\frac{1}{2}}$. The width $\Delta\omega=\frac{\omega_1}{k_1a}-\omega_{\pi}$ of this very wide hyperbola-like resonant band is infinite.


\begin{thebibliography}{99}
\bibitem{ceff} For the $S_0$ branch we calculate $c_{\mathrm{eff}}=3503~\mathrm{m~s^{-1}}$, using the parameters of \tref{table1}.
\end{thebibliography}
\end{document}